\documentclass{aa}  
\usepackage{graphicx}
\usepackage{lscape}
\usepackage{aalongtable,lscape}
\usepackage{rotating}
\usepackage{txfonts}
\begin{document}
 \title{Chemical composition of the young open clusters IC~2602 and IC~2391
\thanks{Based on observations collected at La Silla and Paranal Observatory,
ESO (Chile). Programs 55.E-0808, 66.D-0284.}}

%  \subtitle{}

   \author{V. D'Orazi
          \inst{1,2}
          \and S. Randich
          \inst{2}
          }

   \offprints{V.D'Orazi}

   \institute{Dipartimento di Astronomia e Scienza dello Spazio, Universit\`a di Firenze, Largo E. Fermi 2, Firenze, 
                Italy
              \email{vdorazi@arcetri.astro.it}
         \and
             INAF $-$ Osservatorio Astrofisico di Arcetri, Largo E. Fermi 5, Firenze, Italy\\
             }

   \date{}

% \abstract{}{}{}{}{} 
% 5 {} token are mandatory
 
  \abstract
   {Galactic open clusters have been long recognized as one of the best tools to investigate the chemical content of Galactic disk and its time evolution. 
In the last decade, many efforts have been directed to chemically characterize the old and intermediate age population; surprisingly, 
the chemical content of the younger and close counterpart remains largely undetermined.} 
   {In this paper we present the abundance analysis of a sample of 15 G$/$K members of the young pre-main sequence clusters IC~2602 and IC~2391.
Along with IC~4665, these are 
the first pre-main sequence
clusters for which a detailed abundance determination has been carried out
so far.}
{We analyzed high-resolution, high S/N spectra acquired with different instruments (UVES and CASPEC at ESO, and the echelle spectrograph at CTIO), using 
MOOG and equivalent width measurements. Along with metallicity ([Fe/H]), 
we measured Na~{\sc i}, Si~{\sc i}, Ca~{\sc i}, Ti~{\sc i} and Ti~
{\sc ii}, and Ni~{\sc i} 
abundances. Stars cooler than $\sim 5500$~show lower Ca~{\sc i},
Ti~{\sc i}, and Na~{\sc i} than warmer stars. By determining Ti~{\sc ii} 
abundances, we show that, at least for Ti, this effect is due to NLTE
and over-ionization.}
{We find average
metallicities [Fe/H] =0$\pm 0.01$ and 
[Fe/H]=$-0.01\pm 0.02$ for IC~2602 and IC~2391, respectively. 
All the [X/Fe] ratios show a solar composition; 
the accurate measurements allow us to exclude the presence of 
star-to-star scatter among the
members.} 
   {}
 \keywords{stars: abundances $-$ Galaxy: open clusters and associations: individual: IC~2602, IC~2391}
 \titlerunning{Abundances in IC~2602 and IC~2391}
 \authorrunning{V. D'Orazi \& S. Randich}
 \maketitle
\section{Introduction}\label{intro}
Open clusters (OCs) are widely recognized as excellent tracers of the chemical properties of the Galactic disk and of its evolution with time. 
In recent years, several studies have been devoted to 
the determination of the chemical composition of the older and intermediate
age population, i.e. OCs with ages greater than $\sim$ 500 Myr,
with the main goal of deriving the radial metallicity gradient across
the Galactic disk (e.g., Carraro et al.~2004; Yong et al.~2005; Friel~2006; Randich et al.
~\cite{ran06}; Carretta et al.~\cite{car07};
Sestito et al. 2007, 2008; Pace et al.~\cite{pace}).
On the other hand, 
less attention has been paid to the abundances of younger 
clusters and, in particular, the so-called pre-main sequence (PMS) clusters (age $\sim$10-50 Myr). 
As we will discuss below, knowledge of the abundance
pattern in these clusters  
is instead important to address different astrophysical issues, such
as the evolution of debris disks, the common origin of young open clusters
and associations,
and the chemical evolution of the solar vicinity. 

First, the age interval defined by PMS clusters is critical to trace
the early stages of evolution of
dusty debris disks around main-sequence stars. The fraction of debris
disks in clusters and the level of 24 $\mu$m excess emission rise
for stars with ages between 5-10 Myr, reach a peak around 20-30~
Myr, and then start decaying inversely with age (e.g., Currie et al.
2008). Planet formation models suggest that terrestrial planets
also reach their final mass by 10-30~Myr; hence,
a good characterization
of debris disks in this critical age range 
provides an indirect tool to study planet formation and evolution (e.g Siegler
et al. 2007).
Several studies have focused on the detection and characterization
of debris disks both in the field and in open clusters and 
on the existence of possible correlations
between the frequency of dusty disks and the fundamental properties 
of the parent-star (spectral type, mass, and luminosity for instance). 
Given the metal-rich nature of stars with giant planets (e.g. Santos et al. 
2004), the correlation between
metallicity and frequency of debris disks has been (partially) investigated: 
the lack of such correlation
(Beichman et al. 2006; Bryden et al. 2006; Trilling et al. 2008)
might reflect the different formation histories of giant planets and debris 
disks. On the other hand, the latter studies have focused on old
solar-type stars in the field (ages $\sim$ few Gyr) and the
younger population has not been involved at all: 
there is instead the possibility of an initial correlation 
between dust production and
metallicity around young stars, but this relationship 
might have disappeared as the stars age (Bryden et al. 2006; 
Dominik \& Decin 2003). PMS clusters and their young population 
represent a unique tracer to address this topic. In particular,
the determination of the metallicity of a 
large sample of PMS clusters with available information
on debris disks would allow
investigation of whether a correlation between metallicity and frequency of 
debris disks is indeed present at young ages.

In a different context, 
the identification of small close-by associations 
(within $\sim$ 100 pc from the Sun) 
along with the knowledge of 
several complexes (e.g. the Sco-Cen association and the Tucana/Horologium) a new impulse has been given to the investigation of the so-called 
superclusters.
The concept of such stellar aggregates with parallel space motions and located in the solar surroundings was introduced by Eggen more than thirty years ago;
one of these superclusters is the Local Association (Eggen 1975, 1983a,b): with the Pleiades, other clusters like $\alpha$ Persei,
$\delta$ Lyrae, NGC~1039, NGC~2516, and the Sco-Cen OB association, 
also IC~2602 would be contained within it (Ortega et al. 2007).
The secure knowledge
of the chemical composition and the detection of a common/different
abundance pattern is critical
to confirm or not a common origin, independently of dynamical/kinematical information. 

Finally, young clusters in the solar vicinity
allow us to chemically characterize our 
immediate surroundings in the disk, the solar neighborhood, at the present time;
at variance with old field stars that may have moved from their original
birthplace, these clusters are very
likely born close to where they are now and thus their abundance
pattern should be representative of the present-day composition
of the solar vicinity. Therefore, the 
comparison of the abundance pattern with that of older stars, including the 
Sun, allows us to place constraints on models of Galactic evolution
in the solar neighborhood.

To our knowledge, only one PMS cluster has an available
metallicity determination. Shen et al. (2005) presented an
abundance analysis of 18 F$-$K dwarf members of the 
$\sim$30 Myr old open cluster IC~4665. They obtained
a very close to solar metallicity ([Fe/H]=$-$0.03) 
with no dispersion among the members (standard deviation 0.04~dex), 
within the observational uncertainties. 

With this background in mind, we started a project 
aimed to derive the chemical composition in young
clusters. As part of this effort,
we present here an abundance analysis of G/K-type 
members of the two PMS clusters IC~2602 and IC~2391. 

The paper is organized as follows: in Sect.~\ref{obs} 
we present the target clusters, observations, and the data reduction; the analysis is described in Sect.~\ref{analysis}, while the 
results are presented in Sect.~\ref{results}; the 
scientific implications are discussed in Sect.~\ref{discussion}, 
followed by the conclusions (Sect.~\ref{conclusions}). 

\section{Target clusters, observations, and data reduction}\label{obs}
%
%\subsection{IC~2602 and IC~2391}
%
IC~2602 and IC~2391 are both PMS clusters with estimated ages of
$\sim$ 30 Myr (Stauffer et al. 1997) 
and $\sim$ 55 Myr (Barrado y Navascu\'es et al. 1999, 2004), respectively. 
The age of IC~2602 has been estimated by classical
isochrone fitting,
while the age of IC~2391 has been derived using the lithium depletion boundary
method.

Due to both their age and distance from the
Sun, several studies have focused on these two clusters, 
aiming to determine different fundamental characteristics, 
such as membership, the lithium distribution among G/K-type stars, 
the X-ray properties, 
rotational velocities, chromospheric activity, and initial mass function 
(Randich et al. 1995, 1997, 2001; Stauffer et al. 1989, 1997; 
Prosser et al. 1996; Patten \& Simon 1996; Spezzi
et al.~\cite{spezzi}). More recently,
Siegler et al.~(\cite{sig}) presented 24 $\mu$m {\it Spitzer} 
MIPS observations of the central field of IC~2391, which allowed them to
to determine the fraction of debris disks around 
stars across a wide range of spectral types, from A to K. 
The comparison with field stars and other clusters indicates
that the fraction of disks
among A-type stars is lower than expected.

Surprisingly, no extensive abundance study has been carried out 
for IC~2602 and IC~2391; Randich et al. (2001) derived the iron
content ([Fe/H]) for a small sub-sample of stars in both clusters,
finding close-to-solar metallicities with rather large standard deviation around the average. Our goal here is twofold: first, based on higher 
quality spectra, we will derive
a more accurate [Fe/H] for the two clusters. 
Second, besides iron, we will also determine abundances 
of other elements.
Our sample includes eight and seven members of IC~2602 and IC~2391, 
respectively, for a total of 15 stars.
From the initial and larger sample of confirmed members in
Randich et al. (\cite{ran01}), 
we selected only high-quality spectra of G- and K-type stars with
rotational velocities (v$\sin{i}$) below $\sim$~20 km/sec, 
since the analysis is much less reliable for stars with higher $v\sin{i}$ values.

The observations were carried out at the European Southern Observatory (ESO) 
and at the Cerro-Tololo Inter-American Observatory (CTIO). New observations 
were obtained for 12 stars using UVES on VLT/UT2 
(Dekker et al. 2000) during three nights of observations on 16, 17 and 18
February of 2001;  a slit of 0.8", providing a resolving power of R$\sim$40000, and the CD4
cross-disperser (5770$\AA$ $-$ 9420 $\AA$) were used. Only the blue
part of the spectrum (5750 -- 7450 \AA) was employed for
the analysis, for consistency
with the analysis of the spectra obtained with other spectrographs (see below).
For the remaining three stars we used the spectra already published by
Stauffer et al.~(\cite{stauf2}) and Randich 
et al.~(\cite{ran97},~\cite{ran01}). Specifically, two stars were observed
at CTIO with the 4m telescope in 
conjunction with the Red Long-Camera echelle spectrograph (RLCeS)
and a 31.6 lines mm$^{-1}$. A 120 $\mu$m slit width (0.8 arcsec on the sky) 
and a Tektronix 2048x2048 
CCD provided a resolution of R~$\sim$43800 and a spectral 
coverage between 5800 and 8200 $\AA$. Finally, we also added one star, 
IC2602-R95, whose spectrum was acquired on April 1994 with the
CASPEC echelle spectrograph mounted at the 3.6m ESO telescope.
A 31.6 lines  mm$^{-1}$ grating together with the red cross disperser was used. The slit width was 300 $\mu$m (2.25 arcsec on the sky), 
resulting in a resolving power of 
R$\sim$18 000 and a spectral coverage from 5585 $\AA$ to 8400 $\AA$.

\begin{figure}[htbp]
\includegraphics[angle=270, width=9cm]{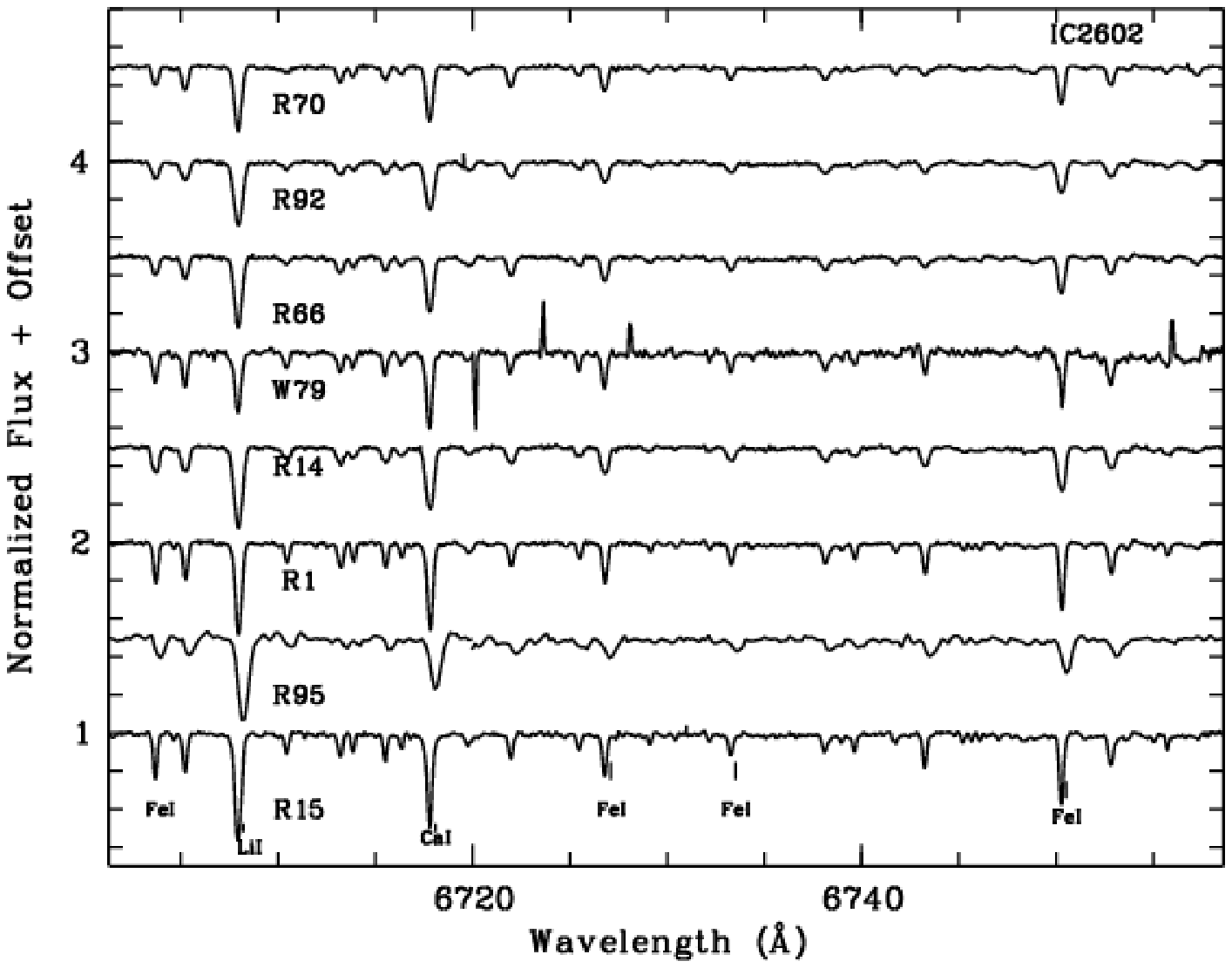}\\ 
\includegraphics[angle=270, width=9cm]{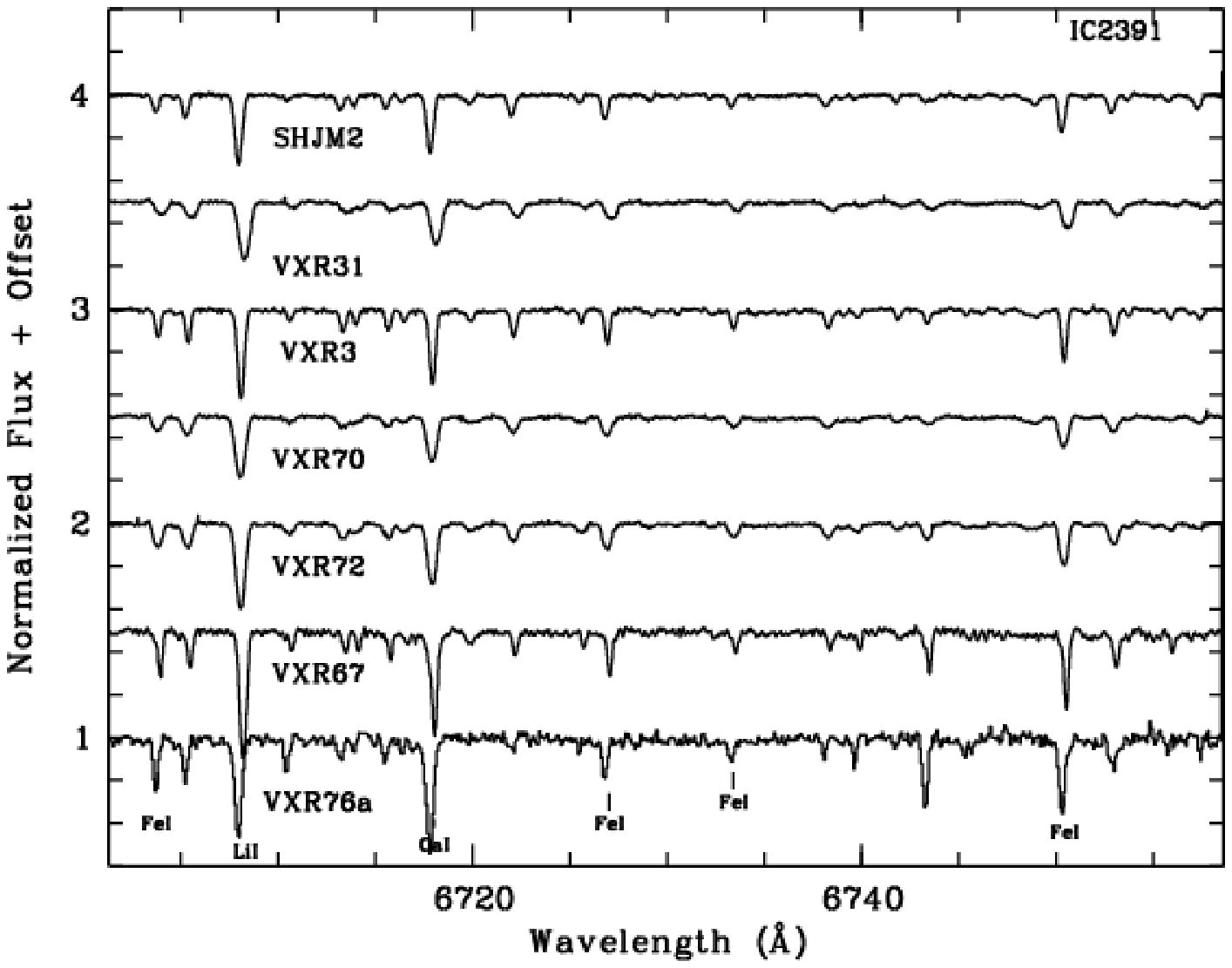}
\caption{Spectra of all the sample stars in a wavelength window of 
60~$\AA$.
Members of IC~2602 and IC~2391 are shown in the upper and lower panels,
respectively.
Several spectral features are marked.}
\label{spectra}
\end{figure}

Information on the target stars is provided in Table~\ref{log}, while
their spectra are shown in Fig.~\ref{spectra} where we plot a 60~$\AA$ 
wide range from 6700 $\AA$ to 6760 $\AA$. 
\setcounter{table}{0}
\begin{table*}
\caption{Sample stars: the first eight 
stars (R1-W79) are members of IC~2602, while the 
remaining seven belong to IC~2391. }
\label{log}
\hspace*{4cm}
\begin{tabular}{cccccccc}
\hline
star & inst. & exp.time & S/N & V & B$-$V & T$_{\rm eff}$ & $v\sin{i}$\\
     &       & (s)    &     &   &     &  (K) &   (km/s)\\
(1)  &  (2)  &  (3)     &  (4)    &  (5)  &  (6)  & (7)  & (8) \\
\hline
R1   & UVES  &  5x870   &  200   & 11.6 & 0.91 & 5050 & $\leq$10\\
R14  & UVES  &  5x870   &  230   & 11.6 & 0.87 & 5150 & 13\\
R15  & UVES  &  10x870  &  200   & 11.7 & 0.93 & 4810 & 7\\
R66  & UVES  &  4x860   &  220   & 11.1 & 0.68 & 5560 & 12\\
R70  & UVES  &  5x900   &  250   & 10.9 & 0.69 & 5700 & 12\\
R92  & UVES  &  7x300   &  200   & 10.3 & 0.69 & 5630 & 14\\ 
R95  & CASPEC & 1200    &  150   & 11.7 & 0.87 & 5020 & 12\\ 
W79  & RLCeS   & 900    &  100   & 11.6 & 0.83 & 5260 & 8\\
     &        &         &      &      &     &    & \\
VXR3 &  UVES & 5x900    &  260     & 10.9 & --- & 5590 & 10\\   
VXR31 & UVES & 7x900    &  280    & 11.2 & --- & 5630 & 17\\
VXR67 & UVES & 8x870    &  130    & 11.7 & --- & 4750 & 8\\
VXR70 & UVES & 5x900    &  260    & 10.8 & 0.64 & 5557 & 17\\ 
VXR72 & UVES & 8x900    &  250    & 11.5 & 0.73 & 5260 & 15\\
VXR76a & RLCeS  & 900    &   70    & 12.8 & 1.05 & 4400 & 8\\
SHJM2 & UVES & 7x300    &  270    & 10.3 & 0.57 & 5970 & $\leq$15\\
\hline
\end{tabular}
\begin{list}{}{}
\begin{footnotesize}
\item (1) Identifier from Randich et al.~(\cite{ran95}) 
for IC~2602 stars with the exception of W79 
(Whiteoak 1961), and from Patten and Simon (\cite{patten}) for IC~2391 
members; (2) instruments (UVES, CASPEC and Red Long-Camera echelle 
spectrograph -RLCeS); (3) exposure time and number of exposures; (4) 
S/N ratios at 6707.78~\AA; (5) V magnitude from Randich et al. (2001); 
(6) B$-$V colors from Randich et al.( 2001); (7) 
initial T$_{\rm eff}$ from Randich et al. (\cite{ran01}),  and (8) v$\sin{i}$
from Stauffer et al. (1997).
\end{footnotesize}
\end{list}
\end{table*}
For details on the reduction of CTIO and CASPEC spectra we refer to 
Stauffer et al. (1997) and Randich et al.~(\cite{ran01}), respectively. 
The UVES data were reduced
using the UVES pipeline Data Reduction Software (Modigliani et al. 2004). 
The software is designed to automatically reduce point source object spectra 
from raw frames to order extraction and final order merging, by following 
the standard procedure: background subtraction, flat-field correction, 
order extraction, sky
subtraction and wavelength calibration.
\section{Analysis}\label{analysis}
\subsection{Line list and equivalent widths}
The abundance analysis was carried out using 
the driver {\it abfind} of the spectral code MOOG (Sneden, 1973 -2000 version)
and a grid of 1-D model atmospheres by Kurucz (1993).
Radiative and Stark broadening are treated in a standard way; for 
the collisional damping we used the classical Uns\"{o}ld (1955) approximation.
As discussed by Paulson et al. (2003) this choice
should not greatly affect the differential analysis with respect
to the Sun. 
LTE abundance values were derived by means of the equivalent widths (EWs)
which were measured using  a Gaussian fitting procedure 
with the IRAF task {\it SPLOT} .

For the analysis we adopted as a basis the line list by Randich et al. 
(2006), but {\bf i)} we selected only lines at $\lambda$ greater 
than 5750~\AA; {\bf ii)} we integrated the list with a few lines
from Sestito et al. (\cite{paola06}), choosing 
the lines included in our wavelength range and 
for which Sestito et al. obtained solar abundances in agreement with the
standard values.
We refer to those two papers for further details on atomic parameters 
($\log~gf$) and their sources; our line list for all the elements with EWs measured both in the solar spectrum
(see next section) and in the spectra of our
sample stars is available as electronic table.
\subsection{Solar analysis}  
Our analysis was carried out differentially with respect to the Sun; 
as a consequence, most of the systematic errors due to e.g. atomic
parameters
should be canceled out. We analyzed the solar spectrum obtained with UVES
(see Randich et al.~2006) 
using the same code, line list, and model atmospheres employed 
for our sample stars. The following stellar parameters were adopted for the 
Sun: T$_{\rm eff}$=5770 K, $\log$g=4.44, 
$\xi$=1.1, and [M/H]=0. We obtained 
$\log$~n(Fe~{\sc I}) = 7.52$\pm$0.02 and $\log$~n(Fe~{\sc II})=7.52 $\pm$0.03.   
The abundance values derived for the other elements are listed in 
Table~\ref{tab2} together with the the values by 
Anders \& Grevesse (\cite{ande}) that
were used as input in MOOG. As the table shows, the agreement
is very good for all elements.
\begin{table}
\caption{Solar abundances derived from our analysis for five elements; 
in the last column we report
the standard values by Anders \& Grevesse (\cite{ande}).}
\label{tab2}
\begin{tabular} {ccc}
\hline
\hline
Element & $\log$~n(X) & $\log$~n(X)$_{\rm AG89}$\\
\hline
Na~{\sc i} & 6.33$\pm$0.02 & 6.33\\
Si~{\sc i} & 7.56$\pm$0.01 & 7.55\\
Ca~{\sc i} & 6.35$\pm$0.02 & 6.36\\
Ti~{\sc i} & 4.97$\pm$0.01 & 4.99\\
Ti~{\sc ii}& 4.92$\pm$0.10 & 4.99\\
Ni~{\sc i} & 6.26$\pm$0.02 & 6.25\\
\hline
\hline
\end{tabular}
\end{table} 
\subsection{Stellar parameters} 
Initial effective temperatures (T$_{\rm eff}$) were retrieved 
from Randich et al. (\cite{ran01}) who, in turn, derived them as using both B$-$V and V$-$I colors and the 
calibrations of Soderblom et al. (\cite{soder})
and Bessell (\cite{bes}), respectively --see Randich et al. (\cite{ran01}) 
for details.

\begin{figure}
\includegraphics[width=9cm]{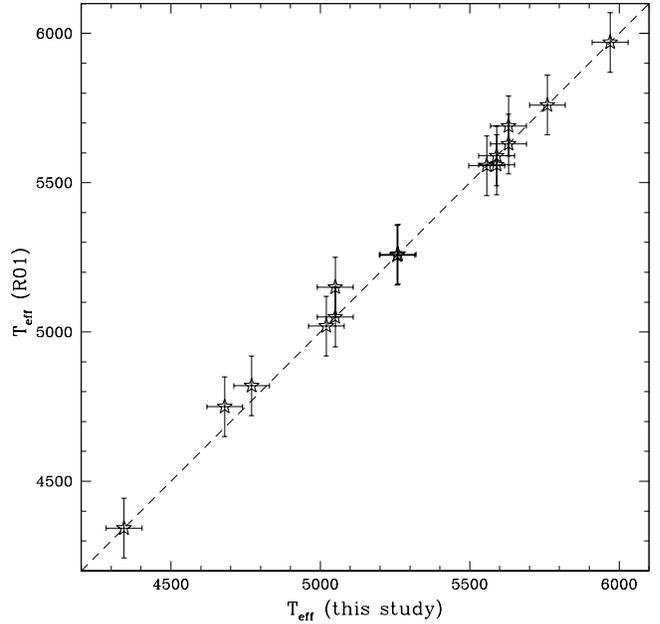} 
\caption{Comparison of the effective temperatures of our sample stars 
with those derived by Randich et al. (\cite{ran01}); the dashed line is the
1:1 relationship.}
\label{teff}
\end{figure}

We formally derived final temperatures by removing the trends between 
$\log$~n(Fe) and the excitation potential ($\chi$), after applying 
a 1-$\sigma$ clipping criterion to the initial line list. 
In most cases, however, initial temperatures were already close to
the final ones and only small adjustments were needed. This is shown
in Fig.~\ref{teff} where we compare our final T$_{\rm eff}$ 
values with those derived by Randich et al. (\cite{ran01}): as one can see,
the agreement is good.

For surface gravities,
we adopted as initial values $\log$g=4.5 for all our sample stars. 
For the warmest stars in the sample
(T$_{\rm eff}> 5550$ K) and for the very slow rotators
(v$\sin{i}< 15$~km/sec) we were able to check
this assumption 
by imposing that the difference between the average $\log$~n(Fe~{\sc I}) and
$\log$~n(Fe~{\sc II}) should be below 0.05 dex.
More in detail, we were able to
measure the Fe~{\sc II} for the stars R66, R70 and R92 (IC~2602) and VXR3
and SHJM2 (IC~2391): the final value that we derived for all these stars is 
$\log$g=4.45 and very agrees well with the initial assumption.

Initial microturbulence values ($\xi$) were estimated using the Nissen's 
relationship (1981): 
$\xi$ = 3.2 x 10$^{-4}$ (T$_{\rm eff}$ $-$ 6390) $-$ 1.3($\log$g $-$ 4.16) 
+ 1.7~km/sec.
Final values were retrieved by zeroing the slope between iron abundances 
and EWs in MOOG.     
\subsection{Errors}       
As usual, the final abundances are affected by 
random (internal) and systematic (external) uncertainties. 

Internal errors due to uncertainties in oscillator strengths ($\log~gf$) 
should cancel out, since our analysis is strictly 
differential with respect to the Sun. 
Random errors due to EW measurements are instead 
well represented by the standard 
deviation from the mean abundance based on the whole set of lines 
($\sigma_1$ in [Fe/H] -see Table~\ref{results}).
The errors ($\sigma_1$) in [X/Fe] listed in Table~\ref{results} were, in turn, obtained by quadratically adding the rms for [Fe/H] 
and rms for [X/H].\\   
Final abundances are also affected by the random errors due to uncertainties in 
the adopted set of stellar parameters, namely T${\rm eff}$, $\log$g, and $\xi$.
These sources of error were estimated by varying one parameter at time, while keeping the other ones unchanged;
then we took the quadratic sum of these three contributions in order to obtain $\sigma_2$, for [Fe/H] and for all [X/Fe] ratios. \\
We adopted random errors of $\pm$60~K and $\pm$0.2~km/sec in T$_{\rm eff}$ 
and $\xi$, respectively: 
variations in T$_{\rm eff}$ larger than 60~K would have introduced
a trend in $\log$~n(Fe) and $\chi$; similarly, variations in microturbulence velocities larger than 0.2 km$s^{-1}$ would result in 
spurious trends in abundances 
versus EWs. 
For the five stars for which we could optimize the surface gravities, the uncertainties in $\log$g were estimated by varying this quantity until the difference between 
$\log$~n(Fe~{\sc i)} and $\log$~n(Fe~{\sc ii)} is larger than 0.05 dex, i.e. the ionization equilibrium condition is no longer verified; this is achieved for
a variation of 0.15 dex (R66 and VXR3) and 0.20 (R70, R92, SHJM2) dex. 
For all the other stars, we adopted a conservative value of 0.25 dex.  
In Table~\ref{errors} we list the errors in [Fe/H] and [X/Fe] ratios due to uncertainties in stellar parameters for one of the coolest stars and for the warmest one in the sample.

The total random errors ($\sigma_{\rm tot}$) were then calculated by quadratically adding $\sigma_1$ and $\sigma_2$; the standard deviation
of the cluster mean, listed in Table \ref{results}, is always smaller 
than the $\sigma_{\rm tot}$ for all our sample stars 
and all the elements, suggesting
that our error estimates are conservative. 

A global estimate of external errors due to the model 
atmospheres and to the code can be obtained by analyzing one star whose chemical content is well known in the literature; to this purpose, we derived abundances for the Hyades member vb187  whose spectrum was acquired with the same instrument (UVES -see Randich et al. 2007). 
The metallicity that we derived ([Fe/H]=0.14$\pm 0.04$) 
is in excellent agreement with all 
previous determinations (e.g. Paulson et al. 2003), 
also [Si/Fe], [Ca/Fe] and [Ti/Fe] well agree within the uncertainties (see Table~\ref{results}). We conclude that our analysis should not be 
affected by major systematic errors.
\begin{table}
\caption{Random errors due to uncertainties in stellar parameters for 
one of the coolest stars and for the warmest one in our sample.}
\label{errors}
\begin{tabular} {lccc} 
\hline
\hline
SHJM2 & T$_{\rm eff}$=5970K & $\log$g=4.45 & $\xi$=1.18~km/s\\
\hline
$\Delta$ & $\Delta$T$_{\rm eff}$=$\pm$60 & $\Delta\log$~g=$\pm$0.25 & $\Delta$$\xi$=$\pm$0.2\\
          &     (K)                       & dex                       & km/s  \\
\hline 	  
[Fe/H]  &     0.05/$-$0.04    & $-$0.03/0.04 & $-$0.02/0.04 \\
${\rm [Na/Fe]}$ & 0.03/$-$0.02 & 0.01/$-$0.01 & 0.02/$-$0.03 \\
${\rm [Si/Fe]}$  &     $-$0.04/0.03     &  0.03/$-$0.03 & 0.03/$-$0.03 \\   
${\rm [Ca/Fe]}$ &     0.03/$-$0.01     & $-$0.03/0.03 & $-$0.02/0.02  \\
${\rm [Ti_I/Fe]}$ &     0.01/$-$0.02     &  0.01/$-$0.01  & 0.01/$-$0.03   \\
${\rm [Ti_{II}/Fe]}$ &    0.02/$-$0.03   & 0.08/$-$0.07 & 0.02/$-$0.01   \\					
${\rm [Ni/Fe]}$ &     $-$0.02/0.01     &  0.02/$-$0.01 & 0.01/$-$0.02  \\
 \hline
         &                &          &         \\
  R95 & T$_{\rm eff}$=5020K & $\log$g=4.5 & $\xi$=0.82 km/s\\
 \hline 
   &                     &             &       \\
$\Delta$ & $\Delta$T$_{\rm eff}$=$\pm$60 & $\Delta\log$~g=$\pm$0.25 & $\Delta\xi=\pm 0.2$\\
                        &     (K)                       & dex                       & km/s \\
 ${\rm [Fe/H]}$  &      0.01/$-$0.02 & $-$0.01/0.01 & $-$0.05/0.04 \\
 ${\rm [Na/Fe]}$ &   0.01/$-$0.03 & $-$0.04/0.01 & 0.05/$-$0.03 \\
 ${\rm [Si/Fe]}$   &	$-$0.03/0.04 & 0.04/$-$0.02 & 0.04/$-$0.03\\ 
 ${\rm [Ca/Fe]}$   &	0.05/$-$0.06 & $-$0.05/0.04 & $-$0.01/0.01\\
 ${\rm [Ti_I/Fe]}$   &    0.07/$-$0.05 &  $-$0.02/0.02 & 0.01/$-$0.02\\
  ${\rm [Ti_{II}/Fe]}$ &    0.05/$-$0.04   &  0.09/$-$0.08  & 0.03/$-$0.04\\
 ${\rm [Ni/Fe]}$  &    $-$0.03/0.02  &  0.05/$-$0.04 & 0.01/$-$0.01\\  	
 \hline 
\end{tabular}
\end{table}
\section{Results}\label{results}
\begin{figure*}
\includegraphics[width=17cm]{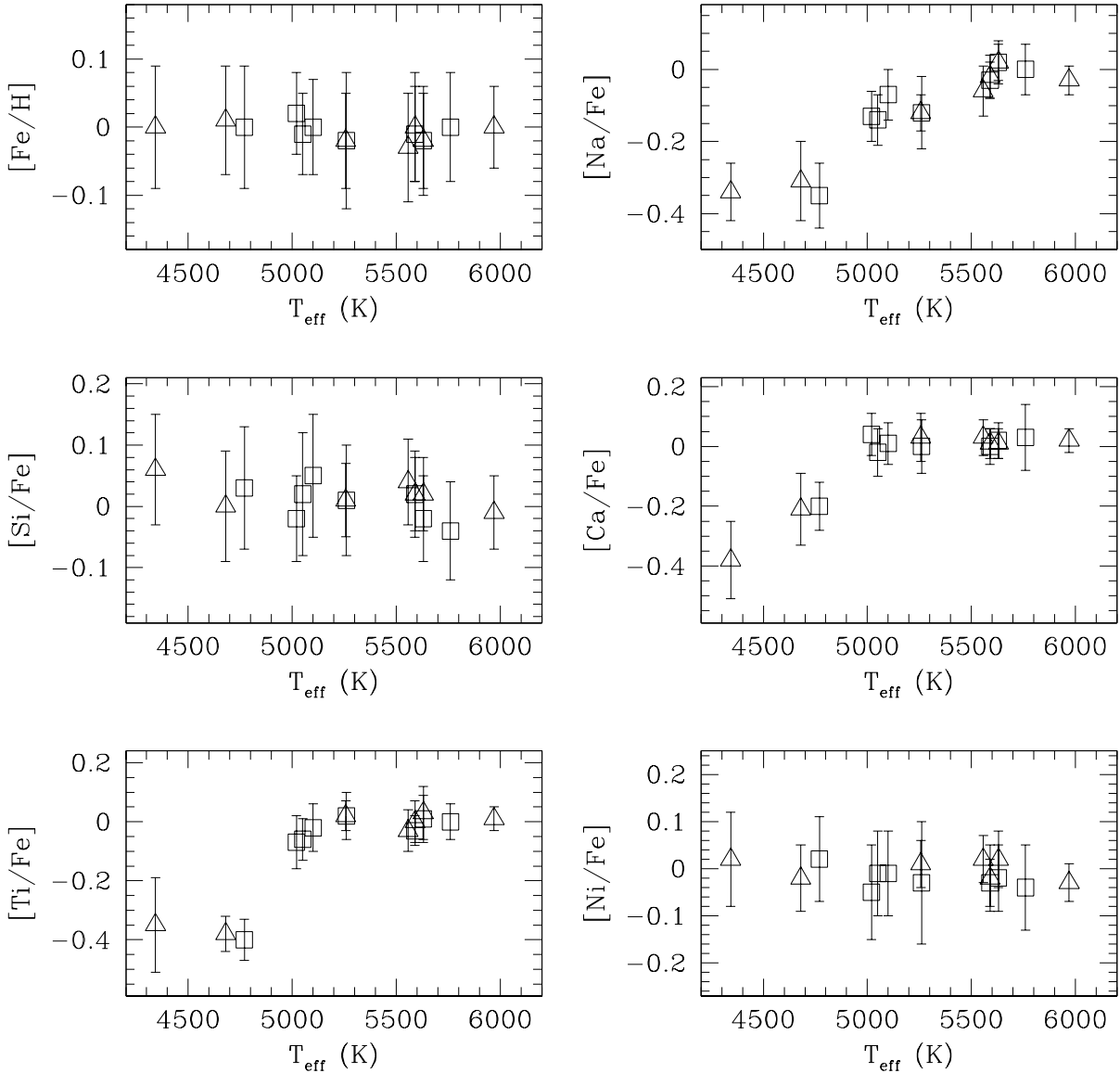} 
\caption{[X/Fe] as function of effective temperatures (T$_{\rm eff}$) for both clusters: squares and triangles are IC~2602 and IC~2391 stars, respectively.}
\label{window}
\end{figure*}
\begin{figure}
\includegraphics[width=8cm]{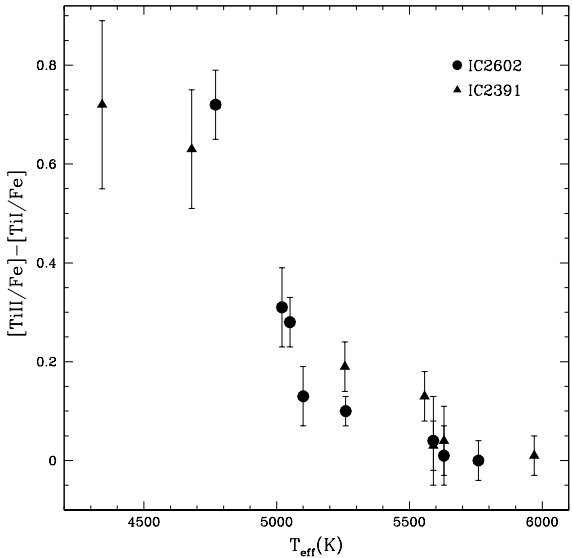} 
\caption{[Ti~{\sc ii}/Fe]-[Ti~{\sc i}/Fe] as a function of T$_{\rm eff}$ 
for both clusters: squares and triangles are IC~2602 and IC~2391 stars, respectively.}
\label{ti2}
\end{figure}
\begin{sidewaystable*}	
\vskip -2.4cm
\caption {Derived abundances. We list again the identifier of each star (1), 
final stellar parameters (Cols. 2, 3, 4), [Fe/H] values (5), and the [X/Fe] 
ratios (Cols.~6, 7, 8, 9, 10). 
The errors ($\sigma_1$, $\sigma_2$) are also reported. Average abundances
for iron and the other elements are also indicated.}
\label{results}
\begin{small}
\begin{tabular} {ccccccccccc} 
\hline
Star & T$_{\rm eff}$ & $\log$g & $\xi$ & [Fe/H] & [Na/Fe] & [Si/Fe] & [Ca/Fe] & [Ti$_{\rm I}$/Fe] & [Ti$_{\rm II}$/Fe] & [Ni/Fe]\\
     &   (K)     &        & (km/sec)     &      &    &           &             &          &   &       \\
     &           &        &                    &     &     &             &        &      &      &  \\
R1   & 5050   & 4.5   & 1.15  & $-$0.01$\pm$0.04$\pm$0.05 & $-$0.14$\pm$0.04$\pm$0.06 & 0.02$\pm$0.05$\pm$0.09 & $-$0.02$\pm$0.07$\pm$0.06 & $-$0.06$\pm$0.05$\pm$0.06 & 
0.22$\pm$0.04$\pm$0.11 & $-$0.01$\pm$0.06$\pm$0.08\\
R14  & 5100 & 4.5 & 1.1 & 0$\pm$0.05$\pm$0.05 & $-$0.07$\pm$0.06$\pm$0.06 & 0.05$\pm$0.06$\pm$0.09 & 0.01$\pm$0.06$\pm$0.06 & $-$0.02$\pm$0.07$\pm$0.06 & 
0.11$\pm$0.04$\pm$0.11 & $-$0.01$\pm$0.08$\pm$0.06\\
R15  &  4770 & 4.5 & 0.7 & 0$\pm$0.06$\pm$0.06 & $-$0.35$\pm$0.07$\pm$0.08 & 0.03$\pm$0.07$\pm$0.09 & $-$0.2$\pm$0.05$\pm$0.07 & $-$0.4$\pm$0.05$\pm$0.06 & 
0.30$\pm$0.05$\pm$0.13 & 0.02$\pm$0.05$\pm$0.08\\
R66 & 5590 & 4.45 & 1.15 & $-$0.01$\pm$0.03$\pm$0.06 & $-$0.03$\pm$0.04$\pm$0.05 & 0.02$\pm$0.05$\pm$0.05 & 0$\pm$0.05$\pm$0.02 & $-$0.03$\pm$0.04$\pm$0.04 & 
0.01$\pm$0.08$\pm$0.09 & $-$0.03$\pm$0.05$\pm$0.04 \\
R70 & 5760 & 4.45 & 1.1 & 0$\pm$0.05$\pm$0.06 & 0$\pm$0.06$\pm$0.05 & $-$0.04$\pm$0.07$\pm$0.06 & 0.03$\pm$0.1$\pm$0.04 & 0$\pm$0.05$\pm$0.03 & 
0$\pm$0.04$\pm$0.08 & $-$0.04$\pm$0.08$\pm$0.05\\
R92 & 5630 & 4.45 & 1.2 & $-$0.02$\pm$0.04$\pm$0.06 & 0.02$\pm$0.04$\pm$0.05 & $-$0.02$\pm$0.06$\pm$0.05 & 0.02$\pm$0.05$\pm$0.03 & 0.01$\pm$0.07$\pm$0.02 & 
0.02$\pm$0.01$\pm$0.08 & $-$0.02$\pm$0.06$\pm$0.04\\
R95 & 5020 & 4.5 & 0.8 & 0.02$\pm$0.03$\pm$0.05 & $-$0.13$\pm$0.05$\pm$0.06 & $-$0.02$\pm$0.06$\pm$0.07 & 0.04$\pm$0.06$\pm$0.07 & $-$0.07$\pm$0.08$\pm$0.07 & 
0.24$\pm$0.04$\pm$0.11 & $-$0.05$\pm$0.09$\pm$0.07\\
W79 & 5260 & 4.5 & 1.1 & $-0.02\pm$0.08$\pm$0.06 & $-0.12\pm$0.09$\pm$0.06 & 0.01$\pm$0.08$\pm$0.07 & 0$\pm$0.08$\pm$0.07 & 0.02$\pm$0.08$\pm$0.05 & 
0.12$\pm$0.01$\pm$0.10 & $-0.03\pm$0.12$\pm$0.04\\
    &      &     &     &                         &       &         &         &                     &                    &\\
IC~2602    &       &    &       &    0$\pm$0.01  & $-0.10\pm$0.12              & 0$\pm$0.03 & $-$0.02$\pm$0.08               & $-$0.06$\pm$0.14 
& 0.13$\pm$0.12 & $-$0.02$\pm$0.02 \\ 
          &      &     &     &                  & $-0.07\pm 0.06^a$ &            &  $0.01\pm0.02^a$  & $-0.01\pm 0.04^a$                &     &             \\
	  &   & & & &  0$\pm 0.02^c$ & & & & &  \\
 &      &     &     &                         &               &         &                     &         &           &\\
VXR3  & 5590 & 4.45 & 1.15 & 0$\pm$0.04$\pm$0.07 & $-$0.02$\pm$0.05$\pm$0.05 & 0.02$\pm$0.06$\pm$0.05 & 0.01$\pm$0.05$\pm$0.02 & 0.00$\pm$0.06$\pm$0.03 & 
0.03$\pm$0.03$\pm$0.09 & $-$0.02$\pm$0.06$\pm$0.04\\
VXR31 & 5630 & 4.5 & 1.2 & $-$0.02$\pm$0.04$\pm$0.07 & 0.02$\pm$0.05$\pm$0.05 & 0.02$\pm$0.05$\pm$0.05 & 0.01$\pm$0.04$\pm$0.01 & 0.03$\pm$0.08$\pm$0.02 & 
0.07$\pm$0.03$\pm$0.09 & 0.02$\pm$0.06$\pm$0.02\\
VXR67 & 4680 & 4.5 & 0.7 & 0.01$\pm$0.05$\pm$0.06 & $-$0.31$\pm$0.09$\pm$0.07 & 0$\pm$0.09$\pm$0.09 & $-$0.21$\pm$0.11$\pm$0.08 & $-$0.38$\pm$0.04$\pm$0.06 & 
0.25$\pm$0.07$\pm$0.13 & $-$0.02$\pm$0.06$\pm$0.06\\
VXR70 & 5557 & 4.5 & 1.15 & $-$0.03$\pm$0.05$\pm$0.07 & $-$0.06$\pm$0.07$\pm$0.05 & 0.04$\pm$0.06$\pm$0.07 & 0.03$\pm$0.06$\pm$0.03 & $-$0.03$\pm$0.07$\pm$0.04 & 
0.09$\pm$0.04$\pm$0.10 & 0.02$\pm$0.05$\pm$0.05\\
VXR72 & 5257 & 4.5 & 1.1 & $-$0.02$\pm$0.03$\pm$0.06 & $-$0.12$\pm$0.05$\pm$0.06 & 0.01$\pm$0.04$\pm$0.08 & 0.03$\pm$0.08$\pm$0.05 & 0.02$\pm$0.05$\pm$0.05 & 
0.17$\pm$0.05$\pm$0.11 & 0.01$\pm$0.04$\pm$0.05\\
VXR76a & 4343 & 4.5 & 1.2 & 0$\pm$0.06$\pm$0.07 & $-$0.34$\pm$0.08$\pm$0.07 &  0.06$\pm$0.09$\pm$0.08 & $-$0.38$\pm$0.12$\pm$0.08 & $-$0.35$\pm$0.16$\pm$0.06 & 
0.37$\pm$0.06$\pm$0.13 & 0.02$\pm$0.09$\pm$0.08 \\
SHJM2 & 5970 & 4.45 & 1.2 & 0$\pm$0.03$\pm$0.05 & $-$0.03$\pm$0.04$\pm$0.04 & $-$0.01$\pm$0.06$\pm$0.06 & 0.02$\pm$0.04$\pm$0.04 & 0.01$\pm$0.04$\pm$0.01 & 
0$\pm$0.02$\pm$0.08 & $-$0.03$\pm$0.04$\pm$0.03\\
 &      &     &     &                         &                        &     &                &           &         &\\
IC~2391  &     &     &       &  $-$0.01$\pm$0.02 & $-$0.12$\pm$0.14              &  0.01$\pm$0.02 & $-$0.07$\pm$0.16  & $-$0.10$\pm$0.18 & 0.14$\pm$0.13 & 0$\pm$0.02\\
       &      &    &       &                 &   $-0.04\pm 0.05^b$ &                & 0.02$\pm 0.01^b$  &  0$\pm 0.02^b$          &        &   \\
       &      &    &       &   & $-0.02\pm 0.03^c$ & & & & & \\
       &      &    &       &                 &               &   &             &    &       &    \\
vb~187 &  5339 & 4.5 & 0.92 &  0.14$\pm$0.04$\pm$0.06 &  0$\pm$0.03$\pm$0.05 & 0.06$\pm$0.02$\pm$0.08 &  0$\pm$0.04$\pm$0.05 &  $-$0.07$\pm$0.04$\pm$0.05 & 0.12$\pm$0.09$\pm$0.09 & 0.04$\pm$0.03$\pm$0.05\\
Hyades$_{\rm Paulson}$ &        &    &      & 0.13$\pm$0.05   & 0.01$\pm$0.09 & 0.05$\pm$0.05   &  0.07$\pm$0.07       &   0.03$\pm$0.05          &  ---  & ---              \\
Hyades$_{\rm Friel}$   &          &         &  &  0.13     & 0.01            & 0.04             & 0.06                &   $-$0.06                  &  --- & --- \\ 
\hline
\end{tabular}
\end{small}
%\begin{footnotesize}
\begin{list}{}{}
\begin{footnotesize}
\item[$^\mathrm{a}$] Average values were computed by discarding the coolest star R15.
 $^b$ VXR67 and VXR76a were not included in the average.
 $^{c}$ Average values by considering only stars with T$_{\rm eff}$ $>$ 5500 K
\end{footnotesize}
\end{list}
%$^a$ Average values were computed by discarding the coolest star R15;
%$^b$ VXR67 and VXR76a were not considered in the mean;
%$^c$ Average values by considering only stars with T$_{\rm eff}$ $>$ 5500 K.
%\end{footnotesize}
\end{sidewaystable*}

Our results for [Fe/H] and all [X/Fe] ratios are reported 
in Table~\ref{results} along with errors due to EW measurements 
and stellar parameters, namely $\sigma_1$ and $\sigma_2$. 
These final abundances were retrieved by applying a 1-$\sigma$ 
clipping to the initial line list for the iron and for the other 
elements with a not 
too small number of lines (such as Ca~{\sc i} and Ti~{\sc i}).
Adopted stellar parameters are also listed in Table~\ref{results}.   

As a verification of our method, we show in Fig.~\ref{window} the [X/Fe] ratios
as a function of T$_{\rm eff}$ for all our sample stars:
as one can see, for [Fe/H], [Si/Fe] and [Ni/Fe] no obvious correlation is 
present in the diagram; on the other hand, sodium, calcium and titanium 
abundances are significantly lower than average for the coolest stars in the 
sample (R15, VXR67, and VXR76a). Trends of [X/Fe] vs. T$_{\rm eff}$
for various elements were already found for old field
stars by Bodaghee et al. (2003) and
by Gilli et al. (2006), who explained them as due
to NLTE effects. Also in our case NLTE is most likely the reason
for the different abundance ratios obtained
for warm and cool stars. However, we note
that, while for Ca we obtain the same decreasing trend 
as in the literature studies, we also find that
[Ti/Fe] ratios decrease with T$_{\rm eff}$, at variance with 
the increasing trend found by both Bodaghee et al. and Gilli et al.
We propose that the different behaviour is due to the much younger age
of our sample stars: they have enhanced levels of chromospheric
activity and are more likely affected by NLTE overionization, rather
than by over-recombination; as discussed by Takeda et al. (\cite{takeda}),
the latter would indeed cause enhancement toward lower 
T$_{\rm eff}$ values. 

To confirm that over-ionization is the reason for
the low Ti~{\sc i} abundances of cool stars, we computed Ti abundances using
a few Ti~{\sc ii} included in our spectral range; the results are shown in
Fig.~\ref{ti2} where [Ti~{\sc ii}/Fe]-[Ti~{\sc i}/Fe] is plotted as a function
of T$_{\rm eff}$. The figure shows that the three stars cooler than
5000~K characterized by lower Ti~{\sc i} abundances indeed have larger differences
between Ti~{\sc ii} and Ti~{\sc i}, confirming the over-ionization hypothesis.

For Na, Mashonkina et al. (2000) computed standard NLTE effects for the lines
used in this paper, finding that for $\log$g$\sim 4.5$, NLTE corrections
are small and do not greatly depend on T$_{\rm eff}$. Again, we suggest
that the low [Na/Fe] ratios that we measure for the coolest sample stars,
might be due to overionization due to chromospheric activity.
We also note that, while Ti and Ca show lower abundances below $\sim 5000$~K,
the trend of decreasing Na already starts at $\sim 5500$~K.
Independently of the explanation for 
the discrepant abundance ratios for cool stars,
we computed the mean [X/Fe] ratios for
Na~{\sc i}, Ca~{\sc i}, and Ti~{\sc i} both considering all stars, and 
discarding the 
three coolest members (see Table~\ref{results}). Average values for Na
were also computed considering only stars warmer than 5500~K. 

Average values of [Fe/H] and [X/Fe] ratios for all elements are shown
in Table~\ref{results}.
The average metallicity is [Fe/H]=0$\pm$0.01
and [Fe/H]=$-$0.01$\pm$0.02 for IC~2602 and IC~2391, respectively; 
the $\alpha$-elements, as well as sodium and nickel, also confirm a 
homogeneous solar composition, with no evidence of star-to-star scatter.
Our results for [Fe/H] are in agreement with
those of Randich et al. (2001) who, as mentioned,
also obtained a solar metallicity for both clusters.
We stress, however, that the higher quality data used here allow us to obtain
better measurements for each star and thus to place tighter constraints on
the average and the absence of any star-to-star variation.
\section{Discussion}\label{discussion}
As mentioned in the Introduction, along with IC~4665, our IC clusters represent the only young PMS clusters for which 
a detailed and accurate chemical composition analysis is available so far.
For both them we obtain 
a solar metallicity and global abundance pattern, similarly
to the results of Shen et al. (2005) for IC~4665. In the following we will
discuss the implications of our results for the three issues discussed
in the introduction.
\subsection{Metallicity and debris disks}
Thanks to the {\it Spitzer Space Telescope}, 
the fraction of debris disks has now been determined in a number of
open clusters and associations
with ages up to 600~Myr (the Hyades). These observations
have allowed investigators to derive the evolution with age of the fraction of debris
disks. As mentioned in the introduction, this number reaches a peak around
20-30~Myr and both the decay and variation of 24~$\mu$m excess ratios
around FGK-type stars appear similar to those measured around A-type stars
(Siegler et al. 2007). The question remains, however, whether this decay
might depend on the metal content.
Siegler et al. (2007) carried out $SPITZER$ 24~$\mu$m observations
of IC~2391 and found that
the fraction of debris disks around stars of spectral type B5-A9
is below that measured in other samples of similar age and even
in older clusters such as the Pleiades and NGC~2516. On the other hand, 
the fraction of debris disks around FGK (solar-type) stars is normal. 
IC~2602 also seems to have a lower than expected fraction of debris
disks around B and A stars. This is shown in Figure~13 of Currie 
et al.~(\cite{currie}),
although no reference is provided for $SPITZER$ observations
of IC~2602.
With the caveat that the lower than expected number of A stars hosting 
debris disks could be simply due to a statistical deviation, 
Siegler et al.~(\cite{sig}) suggested possible environment effects,
among which metallicity, as responsible for this trend. 

To our knowledge, only a minority of the open clusters observed by $SPITZER$
have an accurate metallicity determination available (Pleiades, NGC~2516,
Hyades). Vice versa, several clusters in the age range $\sim 100-600$~Myr
with metallicity measurements do not have $SPITZER$ observations.
In order to reveal the 
possible role of metallicity in debris disk frequency and evolution, a large
sample of clusters with both metallicity and $SPITZER$ measurements should
be constructed; in particular, metallicity should be determined
for additional PMS clusters observed by $SPITZER$.
Nevertheless, the comparison of IC~2391 with the Pleiades and NGC~2516
seems to suggest that the reason for the lower fraction of A-type stars with 24~$\mu$m
excess in IC~2391 is not metallicity. In fact, both NGC~2516
and the Pleiades have a fraction of stars showing an excess 
in agreement with the expected decay law from younger clusters and
field stars, and larger than in IC~2391.
Also, both clusters have a very close to solar metallicity (see
Sect.~\ref{common} below and Terndrup et al. 2002). Our reliable confirmation of a solar
metallicity for IC~2391 (and IC~2602) thus seems to
rule out a metallicity effect.
\subsection{Abundances as tracers of a common origin}\label{common}
We recall that a common origin has been 
suggested for several near-by OCs and young associations like the Pleiades, 
NGC~1039, and IC~2602, among others. In order to check whether
their chemical composition supports the common origin scenario or not,
in Fig.~\ref{origin} 
the abundance pattern of IC~2602 is compared with those of Pleiades
(both the King et al. 2000 and Wilden et al. 2002 analyses are considered), 
and of NGC~1039 (Schuler et al. 2003). Note that
the two studies of the Pleiades yield different metallicities; namely,
[Fe/H]$=0.06\pm0.03$ (King et al. 2000), while Wilden et al. (2002) found 
$\log$~n(Fe)=7.44$\pm 0.02$. This translates into
[Fe/H]=$-0.08$, $-$0.05, and $-0.1$, when assuming
our solar Fe abundance, their solar model Fe abundance ($\log$~n(Fe)=7.49,
their Table~3), or their adopted value
for the Sun ($\log$~n(Fe)=7.54 --their Table~5). We also mention that the
study of Wilden is based on a much large number of stars than that of
King et al.  
Given the uncertainties on the derived abundances, 
the comparison between Pleiades and IC~2602 does not confirm 
nor exclude the presence of a common origin and a
homogeneous analysis is needed. 
On the contrary, the abundance pattern
of NGC~1039 appears to be significantly different from 
that of IC~2602, well beyond the observational uncertainties, not only for
iron, but also for the other elements. In particular, NGC~1039 shows 
an iron content slightly over-solar, with $\Delta$[Fe/H]=+0.07~dex
(Schuler et al. 2003)
with respect to IC~2602; on the other hand, silicon, titanium and nickel seem to show a 
sub solar composition. The [Ca/Fe] ratio is instead close to the solar, although in this case
the uncertainty is much larger, with a standard deviation of $\sim$ 0.1 dex .

To investigate if such a behaviour could be due to 
inhomogeneities in the abundance scales, 
we re-analyzed one member (JP133) of Schuler et al. (2003) using
their published EWs and a) their atomic parameters, b) our atomic
parameters for the lines in common with our list.
Also, we analyzed one of our sample stars (R66), adopting their atomic
parameters for the lines in common.
The results of the different tests are listed in Table~\ref{schuler}: 
both the [Fe/H] values and [X/Fe] ratios
that we obtain for the JP133 are similar to those 
derived by Schuler and collaborators; also, for R66 we retrieve abundances
very similar to our original ones. These results show that the
two abundance scales are consistent with each other.
In other words, we confirm that the observed discrepancy in 
abundance pattern is not related to the abundance analysis, but most likely 
reflects 
the different chemical composition of these two clusters. In turn, these
different abundance patterns do not support a common origin for NGC~1039
and IC~2602.
\begin{table*}
\caption{Re-analysis of a NGC1039 member (JP133) and of R66 (IC~2602). }
\label{schuler}
\begin{tabular}{ccccccccc}
\hline
Star               & T$_{\rm eff}$ & $\log$g & $\xi$ &  [Fe/H]   & [Si/Fe] & [Ca/Fe] & [Ti/Fe] & [Ni/Fe]\\
                   &    (K)        &        & (km/s)        &         &         &         &  \\
                   &               &         &                  &         &            &      &       \\
JP133$_{\rm Schuler}$ &  5710    & 4.54     & 1.88   &  0.07$\pm$0.05 & $-$0.02$\pm$0.08 & 0.02$\pm$0.08 & $-$0.17$\pm$0.08 & $-$0.13$\pm$0.05\\
%JP133$_{\rm 1}$$^\mathrm{*}$  &     5710     &   4.54   &  1.88 &  0.07$\pm$0.06 & $-$0.12$\pm$0.07 & $-$0.14$\pm$0.08 & $-$0.06$\pm$0.03 & -0.17$\pm$0.08 \\
JP133$^a$  &    5710 & 4.54 & 1.88 & 0.08$\pm$0.06   &  $-$0.09$\pm$0.07 & $-$0.02$\pm$0.1 & $-$0.17$\pm$0.11 &
$-$0.16$\pm$0.08\\                  
JP133$^b$ &  5710 & 4.54 & 1.88 & 0.05$\pm$0.14	& --- & $-$0.02$\pm$0.16 & --- & $-$0.18$\pm$0.11\\	
		   &            &           &        &                &                &                &               &             \\
R66                & 5640      &   4.45   &   1.30   &  $-$0.02$\pm$0.07 & $-$0.02$\pm$0.07 & 0.05$\pm$0.06 & $-$0.03$\pm$0.05 & $-$0.04$\pm$0.05\\ 
\hline
\end{tabular}
\begin{list}{}{}
\item[$^\mathrm{a}$ With the line list by Schuler et al.~(2003) and 
our method] 
\item[$^\mathrm{b}$ EWs by Schuler et al. and our $\log gf$ for common lines.] 
\end{list}
\end{table*}
\subsection{IC clusters in the disk}
Based on the fact that the average metallicity of solar-type in the solar
neighborhood is about 0.2~dex below that of the Sun, Wielen et
al. (\cite{wielen}) suggested that the Sun might have been formed
in the inner part of the disk and then migrated to its present position.
The solar metallicity and solar abundance ratios for IC~2602, IC~2391,
and IC~4665 do not support this hypothesis. As mentioned, given their
young age, these clusters have had no time to move from their birthplace,
the solar neighborhood; thus their abundances trace the chemical
composition of the solar vicinity at the present time. The fact that they share
the same composition as the Sun (both iron and other elements) 
provides support to the idea that the Sun was formed
where it is now and, more in general,
that the solar neighborhood underwent very little (if any) chemical
evolution in the last $\sim 4.5$~Gyr.

\begin{figure*}
\includegraphics[width=17cm]{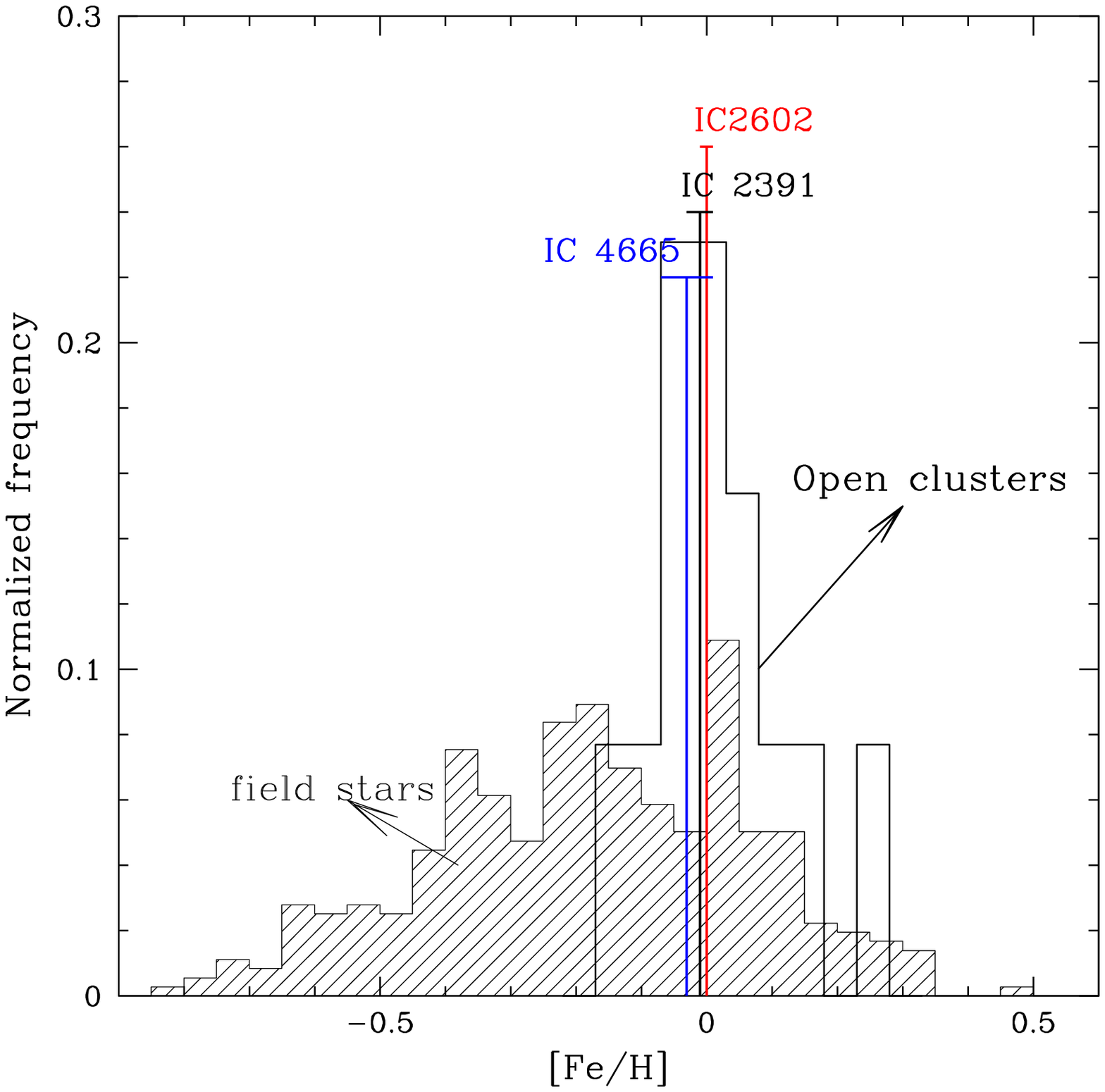}
\caption{Distribution of [Fe/H] of thin disk stars from Soubiran \& Girard
(2005). We include only stars 
in Soubiran \& Girard with available kinematics and consider as thin disk
population stars with P$_{\rm th. disk} \geq 90$\%. This makes a sample
of 358 stars.
The average metallicities of IC~2602, IC~2391, and IC~4665 are also
shown, along with the distribution of [Fe/H] of open clusters within 500~pc
from the Sun with available good measurement of the metallicity.  
References for cluster data are: Ford et al. (2005, Blanco 1), Friel \& Boesgaard (1992, Coma Ber), Paulson et al. 
(2003, Hyades), Wilden et al. (2002, Pleiades), Boesgaard \& Friel (1990, $\alpha$ Persei), Sestito et al. (2003, NGC6475),
Pace et al. (2008, Praesepe), Schuler et al. (2003, NGC1039), Terndrup et al. (2002, NGC2516), H\"unsch et al. (2004,
NGC2451a),  Jeffries et al. (2002, NGC6633), Jacobson et al. (2007, IC4756), Sestito et al. (2004, NGC752). 
 We have considered as thin disk stars those with a probability
of membership to the thin disk $\geq 90$ \%.}
\label{soubiran}
\end{figure*}

\begin{figure*}
\includegraphics[width=17cm]{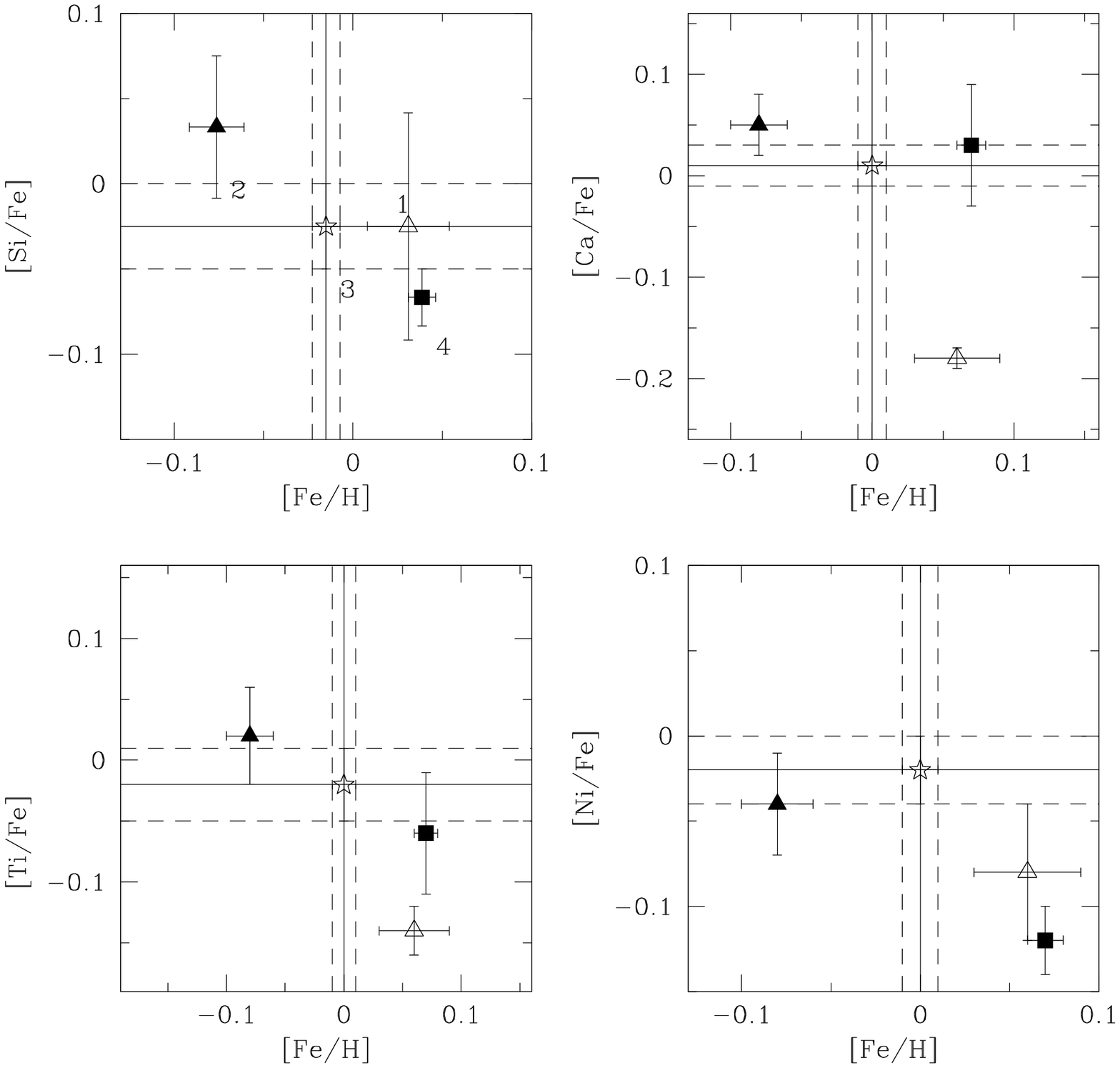}
\caption{Comparison of [X/Fe] ratios as function of [Fe/H] for IC~2602 (our study -star symbol), NGC~1039 (Schuler et al. 2003 -filled square) and 
the Pleiades (Wilden et al. 2002; King et al. 2000 -open and filled triangles.)
The solid lines indicate the average of IC~2602, while dashed lines
indicate average [Fe/H] or [X/Fe] $\pm 1\sigma$.}
\label{origin}
\end{figure*}

In Figure~\ref{soubiran} we compare the metallicity
distribution of thin disk stars from Soubiran \& Girard
(2005) with the average [Fe/H] values
of IC~2602, IC~2391, and IC~4665, as well as 
with the distribution of metallicities
of all clusters closer than 500~pc with a good quality measurement
of [Fe/H]. 
The figure clearly shows that the metallicity
distribution of open clusters is different
from that of field stars; the last extends to significantly
lower metallicities and is characterized by a larger dispersion.
The cluster distribution peaks at the same solar metallicity
of the three PMS clusters; 
the mean metallicity of all the clusters, including the PMS clusters,
is [Fe/H]$_{\rm clust.}=0.02\pm 0.1$, to be compared with
[Fe/H]$_{\rm field}=-0.18 \pm 0.24$ for field stars. The fraction
of field stars with 
a very close-to-solar metallicity ($-0.05 \leq$[Fe/H]$\leq 0.05$) is 14~\% only,
while almost 62~\% have [Fe/H] below
$-0.1$. Also, we
note that the oldest cluster in the sample (NGC~752, $\sim 2$~Gyr) shares
the solar metallicity of the IC~clusters. In other words,
the inclusion of IC~2602 and IC~2391 in the cluster sample 
not only reinforces the conclusion
that most clusters in the solar neighborhood are characterized
by a solar metallicity, but also reveals the lack of
an age-metallicity relationship.

The difference between the young cluster and field stars [Fe/H] values
suggests that a large fraction of field stars
have not formed where they are now in the solar vicinity; rather they 
originated
at larger distances from the Sun, in the outer part of the Galaxy, where
the metallicity is lower because of the presence of a negative
radial metallicity gradient (e.g., Magrini et al. 2009 and references
therein). These
stars then underwent radial mixing (see e.g. Prantzos
2008), moving to their present position.

\section{Summary and conclusions}\label{conclusions}
In this paper, we present a high-resolution spectroscopic study
of the two young near-by PMS clusters IC~2602 and IC~2391;
along with IC~4665, these are the only young OCs whose abundances have 
been determined so far. 
We measured
elemental abundances of Fe~{\sc i}, Na~{\sc i}, Si~{\sc i}, Ca~{\sc i}, 
Ti~{\sc i}, Ti~{\sc ii}, and Ni~{\sc i} for a sample of 15 dwarf members of spectral types 
G$-$K.
The clusters show a highly
homogeneous solar composition, with no hint of dispersion 
among the members, within the very small uncertainties. 

Our results strongly favour the hypothesis that the Sun was born where it
is now, rather than having migrated there from the inner disk, and that the
solar neighborhood has undergone little evolution
in the last 4.5~Gyr. On the other hand,
the comparison with the metallicity distribution of field stars in the solar
vicinity suggests that metal-poor stars in this sample originated
from the outer parts of the Galactic disk.

The comparison between the abundance patterns of IC~2602 and NGC~1039
seems to exclude a common origin for the two clusters, as proposed in the
literature; iron as well the overall abundance pattern are different, 
well beyond the measurement uncertainties. 

Finally, our results seem to exclude metallicity as the reason 
for the low fraction of debris disks observed among A-type stars in IC~2391.
We stress however that, in order to derive definitive
conclusions on the relationship between metallicity and the evolution
of debris disks, accurate metallicity determinations should be performed
for PMS clusters with available $SPITZER$ observations.
\begin{acknowledgements}
This work has made use of the WEBDA database, originally developed by J.C. 
Mermilliod and now maintained by E. Paunzen. We warmly thank the anonymous
referee for her/his valuable comments. 
This project has been partially funded by PRIN INAF 2007 "From active accretion to planetary debris disk: transitional disks and metallicities".
\end{acknowledgements}

\end{document}